\documentclass[notitlepage,nofootinbib,preprintnumbers,amssymb]{revtex4-2}
\usepackage{amsfonts,amssymb,mathtools,graphicx,color,bm}
\definecolor{ultramarine}{rgb}{0.07, 0.04, 0.56}
\definecolor{cadmiumgreen}{rgb}{0.0, 0.42, 0.24}
\definecolor{indigo(dye)}{rgb}{0.0, 0.25, 0.42}
\usepackage[linktocpage=true]{hyperref}
\hypersetup{
colorlinks=true,
citecolor=ultramarine,
linkcolor=cadmiumgreen,
urlcolor=indigo(dye),
}

\usepackage{autobreak}
\newcommand{\relmiddle}[1]{\mathrel{}\middle#1\mathrel{}}
\newcommand{\D}{{\rm d}}
\newcommand{\Ds}{{\rm D}}
\newcommand{\Rs}{{}^{(3)}\!R}
\newcommand{\fr}[2]{\frac{#1}{#2}}
\newcommand{\pa}{\partial}
\newcommand{\ti}{\tilde}
\newcommand{\na}{\nabla}
\newcommand{\bra}[1]{\left( #1 \right)}  
\newcommand{\brb}[1]{\left[ #1 \right]}  
\newcommand{\brc}[1]{\left\{ #1 \right\}}  
\newcommand{\be}{\begin{equation}}  
\newcommand{\ee}{\end{equation}}
\newcommand{\bem}{\begin{bmatrix}}
\newcommand{\eem}{\end{bmatrix}}

\newcommand{\la}{\lambda}
\newcommand{\si}{\sigma}

\newcommand{\mn}{{\mu \nu}}

\newcommand{\mF}{\mathcal{F}}

\newcommand{\mL}{\mathcal{L}}

\newcommand{\mU}{\mathcal{U}}
\newcommand{\mV}{\mathcal{V}}
\newcommand{\mZ}{\mathcal{Z}}

\begin{document}

\preprint{YITP-23-91}

\title{Invertible disformal transformations with arbitrary higher-order derivatives}

\author{Kazufumi Takahashi}
\affiliation{Center for Gravitational Physics and Quantum Information, Yukawa Institute for Theoretical Physics, Kyoto University, 606-8502, Kyoto, Japan}

\begin{abstract}
Invertible disformal transformations serve as a useful tool to explore ghost-free scalar-tensor theories.
In this paper, we construct a generalization of invertible disformal transformations that involves arbitrary higher-order covariant derivatives of the scalar field.
As a result, we obtain a more general class of ghost-free scalar-tensor theories than ever.
Notably, our generalization is such that matter fields can be consistently coupled to these theories without introducing an unwanted extra degree of freedom in the unitary gauge.
\end{abstract}

\maketitle

\section{Introduction}\label{sec:intro}

General relativity (GR) has passed various gravitational experiments as well as cosmological observations and is now commonly accepted as the standard model of gravitation and cosmology.
Nevertheless, there are several motivations to study modifications/extensions of GR.
For instance, GR is expected to be a low-energy effective theory and should be modified at high energies.
Also, extended gravitational theories serve as good candidates that can be tested against GR~\cite{Koyama:2015vza,Ferreira:2019xrr,Arai:2022ilw}.
In general, modified gravity models involve additional degrees of freedom (DOFs) on top of the metric, of which scalar-tensor theories (i.e., those involving a single scalar field besides the metric) have been studied extensively.
The most general class of scalar-tensor theories with second-order Euler-Lagrange equations is now known as the Horndeski class~\cite{Horndeski:1974wa,Deffayet:2011gz,Kobayashi:2011nu}.
Note that the second-order nature of the Euler-Lagrange equations guarantees the absence of Ostrogradsky ghosts, i.e., unstable extra DOFs associated with higher-order equations of motion~\cite{Woodard:2015zca}.

A more general class of ghost-free scalar-tensor theories was constructed in Refs.~\cite{Langlois:2015cwa,Crisostomi:2016czh,BenAchour:2016fzp} by imposing the degeneracy condition~\cite{Motohashi:2014opa,Langlois:2015cwa,Motohashi:2016ftl,Klein:2016aiq,Motohashi:2017eya,Motohashi:2018pxg} on the higher-derivative terms, and this class is called the degenerate higher-order scalar-tensor (DHOST) class.
(See Refs.~\cite{Langlois:2018dxi,Kobayashi:2019hrl} for reviews.)
The DHOST class consists of many subclasses, and one of them can be obtained by the (conformal or) disformal transformation~\cite{Bekenstein:1992pj,Bruneton:2007si,Bettoni:2013diz} of the Horndeski class, which we call the disformal Horndeski (DH) class.
Note in passing that a ghost-free theory is mapped to another ghost-free theory by the disformal transformation since it is invertible in general~\cite{Domenech:2015tca,Takahashi:2017zgr}.
Interestingly, DHOST theories that lie outside the DH class are known to exhibit ghost/gradient instabilities (or otherwise the metric becomes nondynamical) on a cosmological background~\cite{Langlois:2017mxy,Takahashi:2017pje,Langlois:2018jdg}.
Therefore, when one applies the DHOST theories to phenomenology, one usually focuses on the DH class.
It was then realized that the framework of ghost-free scalar-tensor theories can be further extended by requiring the degeneracy only in the unitary gauge where the time coordinate is chosen so that the scalar field is spatially uniform.
Such an extension was dubbed the U-DHOST class~\cite{DeFelice:2018mkq}, which is equivalent to spatially covariant gravity~\cite{Gao:2014soa,Gao:2018znj,Gao:2019lpz,Motohashi:2020wxj} in the unitary gauge.
Note that the scalar field profile has to be timelike in order to be consistent with the unitary gauge.
Away from the unitary gauge, there is an apparent Ostrogradsky mode, but this mode does not propagate as it satisfies a three-dimensional elliptic differential equation on a spacelike hypersurface~\cite{DeFelice:2018mkq,DeFelice:2021hps,DeFelice:2022xvq}.
Such a mode is often called a shadowy mode and is harmless when the scalar field has a timelike profile.

Although the (U-)DHOST theories form a general class of ghost-free scalar-tensor theories, one is interested in further generalizations as they would exhibit peculiar phenomena that could be tested with experiments/observations.
A systematic approach to this issue was proposed in Refs.~\cite{Takahashi:2021ttd,Takahashi:2022mew}.
The idea is to generalize the disformal transformation to incorporate higher-order derivatives of the scalar field, keeping the transformation invertible.
This is indeed possible when the transformation involves (covariant) derivatives of the scalar field up to the second order~\cite{Takahashi:2021ttd}.
(See also Ref.~\cite{Babichev:2021bim} for an earlier attempt in this direction and Refs.~\cite{Domenech:2019syf,Domenech:2023ryc} for a complementary class of invertible disformal transformations with higher derivatives.)
Then, by performing the generalized disformal transformation on the Horndeski theories, one obtains a novel class of ghost-free scalar-tensor theories that goes beyond the conventional DHOST class.
This novel class was dubbed the generalized disformal Horndeski (GDH) theories~\cite{Takahashi:2022mew}.
A further extension can be obtained by choosing the U-DHOST theories as the seed of the generalized disformal transformation, which is called the generalized unitary-degenerate (GDU) theories~\cite{Takahashi:2023jro}.

A possible problem with such generalized disformal theories is that an unwanted extra DOF can show up when matter fields are coupled~\cite{Takahashi:2022mew,Naruko:2022vuh}.\footnote{The problem of matter coupling is absent in the DH theories unless one considers a non-standard matter field whose action contains derivatives of the metric~\cite{Deffayet:2020ypa}.
In this sense, the problem is peculiar to the GDH (or GDU) theories.}
Needless to say, matter fields should be taken into account when we construct gravitational theories that can be used for phenomenological purposes.
In addition, the GDH/GDU theories can be distinguished from the seed Horndeski/U-DHOST theories only in the presence of matter since the matter fields define a special frame where they are minimally coupled to gravity (i.e., the Jordan frame).
Therefore, one is interested in a subclass of the GDH/GDU class where matter fields can be consistently coupled without introducing an unwanted extra DOF.
The consistency of matter coupling in generalized disformal theories has been studied extensively in Refs.~\cite{Takahashi:2022mew,Naruko:2022vuh,Takahashi:2022ctx,Ikeda:2023ntu}, and it was shown that there exists a nontrivial class of generalized disformal theories such that ordinary matter fields, including those in the standard model, can be consistently coupled in the unitary gauge.

Along this line of thought, in the present paper, we construct invertible disformal transformations that involve {\it arbitrary} higher-order covariant derivatives of the scalar field and hence extend the class of generalized disformal transformations obtained in Ref.~\cite{Takahashi:2021ttd}.
These transformations can be employed to obtain a more general class of ghost-free scalar-tensor theories than ever.
Our generalization is such that the resultant theories allow for consistent matter coupling in the unitary gauge.

The rest of this paper is organized as follows.
In \S\ref{sec:2nd}, we briefly review the generalized disformal transformations with second-order covariant derivatives of the scalar field.
We also discuss the conditions on the generalized disformal transformations under which matter fields can be consistently coupled to the GDH/GDU theories.
In \S\ref{sec:higher}, we construct generalized disformal transformations with arbitrary higher-order derivatives of the scalar field in such a way that they respect the conditions for consistent matter coupling.
Finally, we draw our conclusions in \S\ref{sec:conc}.

\section{Transformations with second-order derivatives}\label{sec:2nd}

Let us first provide a brief review of generalized disformal transformations with second-order (covariant) derivatives of the scalar field.
As clarified in Ref.~\cite{Takahashi:2021ttd}, although the inclusion of such higher-order derivatives spoils the invertibility of the transformation in general, one can systematically obtain a class of invertible transformations by focusing on a group structure under functional composition of disformal transformations.
Then, by performing the invertible generalized disformal transformations on Horndeski theories, the authors of Ref.~\cite{Takahashi:2022mew} constructed the class of GDH theories.
Moreover, if one chooses U-DHOST theories as the seed of the generalized disformal transformation, one obtains the class of GDU theories~\cite{Takahashi:2023jro}.
The issue of matter coupling in GDH theories was investigated in Refs.~\cite{Takahashi:2022mew,Naruko:2022vuh,Takahashi:2022ctx,Ikeda:2023ntu}, showing that the generalized disformal transformation is subjected to severe constraints in order to avoid Ostrogradsky ghost in the presence of matter fields.
We note that the analyses in Refs.~\cite{Takahashi:2022mew,Naruko:2022vuh,Takahashi:2022ctx} were performed in the unitary gauge, and hence the scalar field was assumed to have a timelike profile.
Away from the unitary gauge, apparently an extra mode shows up, but it is actually a non-propagating shadowy mode~\cite{DeFelice:2018mkq,DeFelice:2021hps,DeFelice:2022xvq}.
The analysis of Ref.~\cite{Ikeda:2023ntu} shows that matter-coupled GDH theories always have an extra mode, which becomes a shadowy mode when the scalar field profile is timelike.
However, when the scalar field profile is spacelike, the extra mode is nothing but an Ostrogradsky mode, which is problematic.
Therefore, throughout the present paper, we assume that the scalar field has a timelike profile.
We emphasize that there are many situations that allow for a timelike scalar profile, including not only cosmology but also black holes (e.g., Refs.~\cite{Mukohyama:2005rw,Babichev:2013cya,Kobayashi:2014eva,Tretyakova:2017lyg,Chagoya:2018lmv,Motohashi:2019sen,Charmousis:2019vnf,Takahashi:2019oxz,Takahashi:2020hso,Takahashi:2021bml,Mukohyama:2022enj,Mukohyama:2022skk,DeFelice:2022qaz}) and neutron stars (e.g., Refs.~\cite{Babichev:2016jom,Chagoya:2018lmv,Kobayashi:2018xvr,Ikeda:2021skk}).

The subclass of generalized disformal transformations allowing for consistent matter coupling is given by the following form:
    \be
    g_\mn\quad\to\quad
    \bar{g}_\mn[g,\phi]=f_0g_\mn+f_1\phi_\mu\phi_\nu+2f_2\phi_{(\mu}\Ds_{\nu)}X,
    \label{GDT2}
    \ee
where $\phi_\mu\coloneqq \pa_\mu\phi$, $X\coloneqq\phi_\alpha\phi^\alpha$, and $\Ds_\mu$ denotes the covariant derivative on a constant-$\phi$ hypersurface, i.e.,
    \be
    \Ds_\mu \Phi\coloneqq h_\mu^\alpha\na_\alpha \Phi, \qquad
    h_\mu^\alpha\coloneqq \delta^\alpha_\mu-\fr{\phi_\mu\phi^\alpha}{X},
    \label{defD}
    \ee
for any scalar quantity~$\Phi$.
Also, $f_i$'s are functions of $(\phi,X,\mZ)$, with
    \be
    \mZ\coloneqq \Ds_\alpha X\Ds^\alpha X.
    \ee
As pointed out in Ref.~\cite{Takahashi:2021ttd}, a group structure under functional composition of disformal transformations is essential for the invertibility of the transformation:
Once one finds the group structure, it is straightforward to construct the inverse transformation for a given disformal transformation as its inverse element in this group.
In order for the set of transformations of the form~\eqref{GDT2} to have the group structure, the following set of conditions should be satisfied~\cite{Takahashi:2021ttd,Takahashi:2022ctx}:
    \be
    f_0\ne 0, \qquad
    f_0(f_0+Xf_1)-X\mZ f_2^2\ne 0, \qquad
    \bar{X}=\bar{X}(\phi,X), \qquad
    \bar{X}_X\ne 0, \qquad
    \bra{\fr{\mZ}{f_0}}_\mZ\ne 0.
    \label{inv_cond_2nd}
    \ee
Here, we have defined $\bar{X}\coloneqq \bar{g}^\mn\phi_\mu\phi_\nu$, with $\bar{g}^\mn$ being the inverse disformal metric such that $\bar{g}^{\mu\alpha}\bar{g}_{\alpha\nu}=\delta^\mu_\nu$.
Written explicitly,
    \be
    \bar{g}^\mn=\fr{1}{f_0}g^\mn-\fr{1}{f_0(f_0+Xf_1)-X\mZ f_2^2}\brb{\bra{f_1-\fr{\mZ f_2^2}{f_0}}\phi^\mu\phi^\nu+2f_2\phi^{(\mu}\Ds^{\nu)} X-\fr{Xf_2^2}{f_0}\Ds^\mu X\Ds^\nu X}, \qquad
    \ee
and hence
    \be
    \bar{X}=\fr{Xf_0}{f_0(f_0+Xf_1)-X\mZ f_2^2}, \label{barX2}
    \ee
which is a function of $(\phi,X,\mZ)$ in general.
The condition~$\bar{X}=\bar{X}(\phi,X)$ in Eq.~\eqref{inv_cond_2nd} requires that this $\bar{X}$ should be independent of $\mZ$.
Actually, without this condition, a functional composition of disformal transformations yields higher-order derivatives that are not involved in the original transformation law, which spoils the group structure.

It should be noted that the transformation~\eqref{GDT2} satisfies the following two properties:
\begin{enumerate}
\renewcommand{\theenumi}{\Alph{enumi}}
\renewcommand{\labelenumi}{[\theenumi]}
\item \label{propertyA} All the higher-order derivatives of the scalar field appear only through $\Ds_\mu$.
\item \label{propertyB} Projected onto a constant-$\phi$ hypersurface, all the non-conformal terms vanish, i.e., $h_\mu^\alpha h_\nu^\beta \bar{g}_{\alpha\beta}=f_0h_\mn$.
\end{enumerate}
We note that Eq.~\eqref{GDT2} is the most general transformation law with these properties constructed out of $g_\mn$, $\phi_\mu$, and $\pa_\mu X$.
In fact, the properties~[\ref{propertyA}] and [\ref{propertyB}] ensure the consistency of matter coupling in GDH/GDU theories.

To see this, let us consider matter fields that are minimally coupled to gravity, where the gravitational action is given by some invertible generalized disformal transformation of the following action:
    \be
    S_{\rm g}[g,\phi]=\int\D^4x\sqrt{-g}\,\mL(g_\mn,\phi,\phi_\mu,K_\mn,a_\mu,\Rs_{\mu\nu\lambda\sigma}),
    \label{S_g-3D}
    \ee
where $K_\mn$, $a_\mu$, and $\Rs_{\mu\nu\lambda\sigma}$ are the extrinsic curvature, the acceleration vector, and the three-dimensional Riemann tensor associated with a constant-$\phi$ hypersurface, respectively.
It should be noted that the action~\eqref{S_g-3D} does not yield any higher-order time derivatives at least under the unitary gauge and hence does not contain unwanted extra DOFs in itself.
Therefore, the action~\eqref{S_g-3D} encompasses the Horndeski and (a subclass of) U-DHOST theories, and its generalized disformal transformation encompasses the GDH/GDU theories.
When one considers matter fields that are minimally coupled to such generalized disformal theories, it is practically more useful to work in the frame where the gravitational action is described by Eq.~\eqref{S_g-3D}.
Schematically, we consider the following action:
    \be
    S[g,\phi,\Psi]=S_{\rm g}[g,\phi]+S_{\rm m}[\bar{g},\Psi],
    \label{matter-coupled-theory}
    \ee
where $S_{\rm m}$ denotes the matter action and the matter fields are collectively denoted by $\Psi$.
In this frame, there are no higher-order time derivatives in the gravitational action, but the matter coupling could yield Ostrogradsky ghost(s) through the higher derivatives contained in $\bar{g}_\mn$.
The reason why the properties~[\ref{propertyA}] and [\ref{propertyB}] remove the ghost(s) can be understood by taking the unitary gauge and expressing the action in terms of the Arnowitt-Deser-Misner (ADM) variables, i.e., the lapse function~$N$, the shift vector~$N^i$, and the spatial metric~$h_{ij}$.
Written explicitly, the spacetime metric is decomposed as
    \be
    g_\mn \D x^\mu \D x^\nu=-N^2\D t^2+h_{ij}(\D x^i+N^i\D t)(\D x^j+N^j\D t), \label{ADM}
    \ee
and hence $X=-\dot{\phi}^2/N^2$ in the unitary gauge, with a dot denoting the time derivative.
Note that the gravitational action~\eqref{S_g-3D} does not involve the time derivative of $N$ and $N^i$, and therefore these variables are nondynamical when matter fields are absent.
On the other hand, if one introduces matter fields that are coupled to the generalized disformal metric, $\dot{N}$ can show up in the matter sector, which could make $N$ dynamical.
It is precisely the property~[\ref{propertyA}] that allows us to avoid this problem:
Since $\Ds_\mu$ is nothing but the spatial covariant derivative under the unitary gauge, the transformation law~\eqref{GDT2} involves only the spatial derivative of $X$ and hence only the spatial derivative of $N$.
Therefore, for ordinary bosonic matter fields whose action does not involve derivatives of the metric, the property~[\ref{propertyA}] ensures the consistency of matter coupling.
For fermionic matter fields, the situation is more nontrivial as the covariant derivative acting on a fermionic field contains derivatives of the metric (or the tetrad).
Nevertheless, as clarified in Ref.~\cite{Takahashi:2022ctx}, the unwanted $\dot{N}$ can be absorbed into a redefinition of the fermionic field when the spatial metric~$h_{ij}$ is transformed conformally, which is guaranteed by the property~[\ref{propertyB}].\footnote{The redefinition of a fermionic field~$\psi$ is a simple scaling~$\psi\to f_0^{3/2}\psi$~\cite{Takahashi:2022ctx}, and hence the discussion can be extended to the case with multiple fermions.}
Thus, the properties~[\ref{propertyA}] and [\ref{propertyB}] are crucial for the consistency of matter coupling in generalized disformal theories.
(For a more detailed discussion, see Ref.~\cite{Takahashi:2022ctx}.)

\section{Transformations with arbitrary higher-order derivatives}\label{sec:higher}

\subsection{Transformation law}\label{ssec:trnsf_law}

Let us now extend the transformation law~\eqref{GDT2} to include third- or higher-order covariant derivatives of the scalar field, keeping the properties~[\ref{propertyA}] and [\ref{propertyB}].
We assume that derivatives of the metric appear in the transformation law only through the Christoffel symbol contained in the covariant derivative of the scalar field as in Eq.~\eqref{GDT2}.
For instance, when we include third-order derivatives, we introduce a new building block~$\Ds_\mu\mZ$ and consider the transformation law
    \be
    g_\mn\quad\to\quad
    \bar{g}_\mn[g,\phi]=f_0g_\mn+f_1\phi_\mu\phi_\nu+2f_2\phi_{(\mu}\Ds_{\nu)}X+2f_3\phi_{(\mu}\Ds_{\nu)}\mZ,
    \label{GDT3}
    \ee
where $f_i$'s are now functions of $(\phi,X,\mZ,\mU,\mV)$, with
    \be
    \mU\coloneqq \Ds_\alpha X\Ds^\alpha\mZ, \qquad
    \mV\coloneqq \Ds_\alpha\mZ\Ds^\alpha\mZ.
    \ee
This is the most general transformation law with the properties~[\ref{propertyA}] and [\ref{propertyB}] constructed out of $g_\mn$, $\phi_\mu$, and $\pa_\mu \chi$, with $\chi\in \brc{X,\mZ}\eqqcolon E_2$.
Here, the set~$E_2$ consists of scalar quantities that contain up to $\pa^2\phi$ and do not involve $\dot{N}$ under the unitary gauge where $\phi=\phi(t)$.
Note that terms like $\Ds_{(\mu}X\Ds_{\nu)}\mZ$ spoil the property~[\ref{propertyB}] and hence are not included in the transformation law.
Also, we assume that the derivative operator~$\Ds_\mu$ always acts on a scalar quantity, as otherwise one cannot apply the general strategy to construct invertible disformal transformations developed in Ref.~\cite{Takahashi:2021ttd}.

Likewise, it is straightforward to construct the transformation law with $\pa^d\phi$ for $d\ge 3$ in an inductive manner.
Suppose we know the set of scalar quantities~$E_n$ that contain up to $\pa^{n}\phi$ and do not involve $\dot{N}$ under the unitary gauge.
Then, we have
    \begin{align}
    E_{n+1}&=E_n\sqcup
    \brc{\Ds_\alpha \chi \Ds^\alpha \chi' \relmiddle| \chi\in E_n,\chi'\in E_n\backslash E_{n-1}} \nonumber \\
    &=\brc{X}\sqcup\brc{\Ds_\alpha \chi \Ds^\alpha \chi' \relmiddle| \chi,\chi'\in E_n}.
    \end{align}
Here, we define $E_0\coloneqq \emptyset$ and $E_1\coloneqq \brc{X}$, so that we recover $E_2=\brc{X,\mZ}$ and obtain $E_3$ and $E_4$ as
    \be
    \begin{split}
    E_3&=E_2\sqcup \brc{\mU,\mV}, \\
    E_4&=E_3\sqcup \brc{\Ds_\alpha X\Ds^\alpha \mU,\Ds_\alpha \mZ\Ds^\alpha \mU,\Ds_\alpha \mU\Ds^\alpha \mU,\Ds_\alpha X\Ds^\alpha \mV,\Ds_\alpha \mZ\Ds^\alpha \mV,\Ds_\alpha \mU\Ds^\alpha \mV,\Ds_\alpha \mV\Ds^\alpha \mV}.
    \end{split}
    \ee
The number of elements of $E_n$ satisfies the following recurrence relation:
    \be
    |E_0|=0, \qquad
    |E_{n+1}|=\fr{|E_n|\bra{|E_n|+1}}{2}+1 \quad (n\ge 0).
    \ee
For instance, 
    \be
    |E_1|=1, \qquad
    |E_2|=2, \qquad
    |E_3|=4, \qquad
    |E_4|=11, \qquad
    |E_5|=67, \qquad
    |E_6|=2279, \qquad
    \cdots.
    \ee
The large-$n$ behavior of $|E_n|$ is given by $|E_n|\sim 2\times c^{2^n}$, with $c\approx 1.116253032687330$.

Let us now denote the elements of $E_n\backslash E_{n-1}$ by $\chi_n^{(A)}$, with $A=1,2,\cdots,|E_n|-|E_{n-1}|$.
Note that $\chi_n^{(A)}$ contains only $n$th- and lower-order derivatives of $\phi$.
For instance,
    \be
    \chi_1^{(1)}=X, \qquad
    \chi_2^{(1)}=\mZ, \qquad
    \chi_3^{(1)}=\mU, \qquad
    \chi_3^{(2)}=\mV, \qquad
    \chi_4^{(1)}=\Ds_\alpha X\Ds^\alpha \mU, \qquad
    \chi_4^{(2)}=\Ds_\alpha \mZ\Ds^\alpha \mU, \qquad
    \cdots.
    \ee
With this notation, one can write down the most general transformation law with the properties~[\ref{propertyA}] and [\ref{propertyB}] constructed out of $g_\mn$, $\phi_\mu$, and $\pa_\mu \chi$, with $\chi\in E_{d-1}$.
As a result, we obtain the following generalized disformal transformation with $\pa^d\phi$ that accommodates consistent matter coupling:
    \be
    g_\mn\quad\to\quad
    \bar{g}_\mn[g,\phi]=f_0g_\mn+f_1\phi_\mu\phi_\nu+2\phi_{(\mu}\xi_{\nu)}, \qquad
    \xi_\mu\coloneqq \sum_{n=1}^{d-1}\sum_{A}f_{n+1}^{(A)}\Ds_\mu\chi_{n}^{(A)},
    \label{GDTd}
    \ee
where $f_0$, $f_1$, and $f_{n+1}^{(A)}$ are now functions of the elements of $E_{d}$ as well as $\phi$.
Here, the summation over $A$ runs from $1$ to $|E_n|-|E_{n-1}|$ for each $n$.
Therefore, the number of coefficient functions is given by $|E_{d-1}|+2$.
Note that Eqs.~\eqref{GDT2} and \eqref{GDT3} are recovered by setting $d=2$ and $d=3$, respectively, with the identifications~$f_2^{(1)}=f_2$ and $f_3^{(1)}=f_3$.

\subsection{Invertibility condition}\label{ssec:inv_cond}

In what follows, we clarify the condition under which the transformation~\eqref{GDTd} is invertible, i.e., Eq.~\eqref{GDTd} can be solved uniquely for the unbarred metric at least locally in the configuration space.
For this purpose, we first need to know the inverse metric associated with the disformal metric~$\bar{g}_\mn$.
As explained in Ref.~\cite{Takahashi:2021ttd} for the case of $d=2$ [i.e., for transformations of the form~\eqref{GDT2}], it is straightforward to find the inverse metric~$\bar{g}^\mn$ for arbitrary $d$.
In the present case, we obtain
    \be
    \bar{g}^\mn
    =\fr{1}{f_0}\bra{g^\mn+\fr{f_0^2-\mF}{X\mF}\phi^\mu\phi^\nu-\fr{2f_0}{\mF}\phi^{(\mu}\xi^{\nu)}+\fr{X}{\mF}\xi^\mu\xi^\nu},
    \label{inverse_metric}
    \ee
where we have defined
    \be
    \mF\coloneqq f_0(f_0+Xf_1)-X\xi_\alpha\xi^\alpha.
    \label{calF}
    \ee
Indeed, the above $\bar{g}^\mn$ satisfies $\bar{g}^{\mu\alpha}\bar{g}_{\alpha\nu}=\delta^\mu_\nu$.
Note that we assume $f_0\ne 0$ and $\mF\ne 0$.
The inverse disformal metric can be used to construct the barred counterparts of scalar quantities~$\chi_n^{(A)}$.
For instance, the barred counterparts of $\chi_1^{(1)}=X$ and $\chi_2^{(1)}=\mZ$ are given by
    \be
    \bar{X}\coloneqq \bar{g}^{\alpha\beta}\phi_\alpha\phi_\beta
    =\fr{Xf_0}{\mF}, \qquad
    \bar{\mZ}\coloneqq \bar{g}^{\alpha\beta}\bar{\Ds}_\alpha \bar{X}\bar{\Ds}_\beta \bar{X},
    \label{barXZ}
    \ee
where $\bar{\Ds}_\mu$ is related to the unbarred derivative operator~$\Ds_\mu$ by
    \be
    \bar{\Ds}_\mu\Phi
    \coloneqq \bra{\delta^\alpha_\mu-\fr{\phi_\mu\phi_\nu}{\bar{X}}\bar{g}^{\nu\alpha}}\pa_\alpha\Phi
    =\Ds_\mu\Phi+\fr{1}{f_0}\bra{\xi^{\alpha}\Ds_\alpha\Phi}\phi_\mu,
    \label{barD}
    \ee
with $\Phi$ being any scalar quantity.
One can show that
    \be
    \bar{g}^{\alpha\beta}\bar{\Ds}_\alpha\Phi\bar{\Ds}_\beta\Phi'
    =\fr{1}{f_0}\Ds_\alpha\Phi\Ds^\alpha\Phi',
    \ee
for arbitrary scalar quantities~$\Phi$ and $\Phi'$.
Note that $\bar{X}$ defined in Eq.~\eqref{barXZ} is a function of the elements of $E_d$ as well as $\phi$ in general.
This means that $\bar{\mZ}$ yields $\pa^{d+1}\phi$, which is not contained in the original transformation law~\eqref{GDTd}.
Likewise, even higher-order derivatives of $\phi$ show up in the barred counterparts of $\chi_n^{(A)}$ in general, which makes it difficult to construct the inverse disformal transformation (see Ref.~\cite{Takahashi:2021ttd} for a more detailed discussion).

In order for the transformation to be invertible, we require that $\bar{\chi}_n^{(A)}$ ($n=1,2,\cdots,d$) is a function of $\phi$ and the elements of $E_n$ only, or equivalently,
    \be
    \fr{\pa\bar{\chi}_n^{(A)}}{\pa\chi_m^{(B)}}=0 \quad (1\le n<m\le d).
    \label{inv_cond1}
    \ee
For instance, $\bar{X}=\bar{X}(\phi,X)$ and $\bar{\mZ}=\bar{\mZ}(\phi,X,\mZ)$ follow from the above condition with $n=1$ and $2$, respectively.
We note that the functional form of $\bar{\chi}_n^{(A)}$ is determined by the coefficient functions in the transformation law~\eqref{GDTd}, and therefore the condition~\eqref{inv_cond1} imposes a set of constraints on those coefficient functions.
Note also that this condition ensures the closedness of the functional composition of two disformal transformations, which allows us to construct the inverse disformal transformation systematically~\cite{Takahashi:2021ttd}.
On top of Eq.~\eqref{inv_cond1}, we require the following condition:
    \be
    \left|\fr{\pa\bar{\chi}_n^{(A)}}{\pa\chi_n^{(B)}}\right|\ne 0 \quad (n=1,2,\cdots,d).
    \label{inv_cond2}
    \ee
This condition allows us to express $\chi_n^{(B)}$ as a function of $\bar{\chi}_m^{(A)}$ with $m\le n$.
For example, the above condition implies $\bar{X}_X\ne 0$ and $\bar{\mZ}_\mZ\ne 0$, and therefore we have $X=X(\phi,\bar{X})$ and $\mZ=(\phi,\bar{X},\bar{\mZ})$.

Surprisingly, one can show that the conditions~\eqref{inv_cond1} and \eqref{inv_cond2} are equivalent to the following simpler condition:
    \be
    \bar{X}=\bar{X}(\phi,X), \qquad
    \bar{X}_X\ne 0, \qquad
    \bar{\mZ}=\bar{\mZ}(\phi,X,\mZ), \qquad
    \bar{\mZ}_\mZ\ne 0,
    \label{inv_cond}
    \ee
which is nothing but the $n=1$ and $n=2$ parts of Eqs.~\eqref{inv_cond1} and \eqref{inv_cond2}.
In other words, the remaining parts of the conditions~\eqref{inv_cond1} and \eqref{inv_cond2} (i.e., $3\le n\le d$) are redundant.
This can be verified by explicit computation of $\bar{\chi}_n^{(A)}$.
For instance, under the condition~\eqref{inv_cond}, we obtain
    \be
    \bar{\mZ}=\fr{\bar{X}_X^2}{f_0}\mZ, \qquad
    \bar{\mU}=\fr{\bar{X}_X}{f_0}\bra{\bar{\mZ}_\mZ\mU+\bar{\mZ}_X\mZ}, \qquad
    \bar{\mV}=\fr{1}{f_0}\bra{\bar{\mZ}_\mZ^2\mV+2\bar{\mZ}_X\bar{\mZ}_\mZ\mU+\bar{\mZ}_X^2\mZ}.
    \ee
The first equation together with Eq.~\eqref{inv_cond} implies $f_0=f_0(\phi,X,\mZ)$, so that we have $\bar{\mU}=\bar{\mU}(\phi,X,\mZ,\mU)$, $\bar{\mV}=\bar{\mV}(\phi,X,\mZ,\mU,\mV)$, and
    \be
    \left|\fr{\pa(\bar{\mU},\bar{\mV})}{\pa(\mU,\mV)}\right|
    =\fr{\bar{X}_X\bar{\mZ}_\mZ^3}{f_0^2}\ne 0.
    \ee
Therefore, we see that the conditions~\eqref{inv_cond1} and \eqref{inv_cond2} are satisfied for $n=3$ (recall that $\chi_3^{(1)}=\mU$ and $\chi_3^{(2)}=\mV$).
Likewise, one can straightforwardly recover the parts of Eqs.~\eqref{inv_cond1} and \eqref{inv_cond2} with $3\le n\le d$ from the condition~\eqref{inv_cond}.

Let us now specify the independent functional DOFs that characterize the invertible subclass of the generalized disformal transformation~\eqref{GDTd}.
Equation~\eqref{inv_cond} motivates us to regard $\bar{X}=\bar{X}(\phi,X)$ and $\bar{\mZ}=\bar{\mZ}(\phi,X,\mZ)$ (such that $\bar{X}_X\ne 0$ and $\bar{\mZ}_\mZ\ne 0$) as given functions.
Then, from $\bar{X}=Xf_0/\mF$ [with $\mF$ defined in Eq.~\eqref{calF}] and $\bar{\mZ}=\bar{X}_X^2\mZ/f_0$, one can express $f_0$ and $f_1$ in terms of $\bar{X}$ and $\bar{\mZ}$ as well as the other coefficient functions~$f_n^{(A)}$ ($n=2,\cdots,d$) as
    \be
    f_0=\fr{\bar{X}_X^2(\phi,X)\mZ}{\bar{\mZ}(\phi,X,\mZ)}, \qquad
    f_1=\fr{1}{\bar{X}(\phi,X)}-\fr{f_0}{X}+\fr{1}{f_0}\xi_\alpha\xi^\alpha,
    \label{f0f1}
    \ee
where $f_n^{(A)}$'s are encapsulated in $\xi_\mu$ [see Eq.~\eqref{GDTd}].
Note that we have $\mF=X\bar{X}_X^2\mZ/(\bar{X}\bar{\mZ})$.
Thus, the independent functional DOFs are $\bar{X}(\phi,X)$, $\bar{\mZ}(\phi,X,\mZ)$, and $f_n^{(A)}$ ($n\ge 2$), with $f_n^{(A)}$'s being arbitrary functions of $\phi$ and the elements of $E_d$.\footnote{A simple example of invertible generalized disformal transformations of the form~\eqref{GDTd} can be obtained by putting $\bar{X}=X$ and $\bar{\mZ}=\mZ$. In this case, we have $\bar{g}_\mn=g_\mn+(\xi_\alpha\xi^\alpha)\phi_\mu\phi_\nu+2\phi_{(\mu}\xi_{\nu)}$, for which $\bar{\chi}_n^{(A)}=\chi_n^{(A)}$.}

Having obtained the invertibility condition, let us explicitly construct the inverse transformation for the generalized disformal transformation~\eqref{GDTd}.
To this end, one has to know how the building blocks of the transformation are related to barred quantities.
Since $\chi_n^{(A)}$ is a function of $\bar{\chi}_m^{(B)}$ with $m\le n$, the derivative of $\chi_n^{(A)}$ can be written in the form
    \be
    \Ds_\mu\chi_n^{(A)}=\sum_{m=1}^{n}\sum_{B}J_{n m}^{(AB)}\Ds_\mu\bar{\chi}_m^{(B)}, \qquad
    J_{n m}^{(AB)}\coloneqq \fr{\pa\chi_n^{(A)}}{\pa\bar{\chi}_m^{(B)}},
    \ee
which enables us to express $\xi_\mu$ as
    \begin{align}
    \xi_\mu&=\sum_{n=1}^{d-1}\sum_{m=1}^{n}\sum_{A,B}f_{n+1}^{(A)}J_{n m}^{(AB)}\Ds_\mu\bar{\chi}_m^{(B)} \nonumber \\
    &=\Xi\phi_\mu+\sum_{n=1}^{d-1}\sum_{m=1}^{n}\sum_{A,B}f_{n+1}^{(A)}J_{n m}^{(AB)}\bar{\Ds}_\mu\bar{\chi}_m^{(B)}.
    \end{align}
Here, we have employed Eq.~\eqref{barD} and defined the following quantity:
    \be
    \Xi\coloneqq -\fr{1}{f_0}\sum_{n=1}^{d-1}\sum_{m=1}^{n}\sum_{A,B}f_{n+1}^{(A)}J_{n m}^{(AB)}\xi^\alpha\Ds_\alpha\bar{\chi}_m^{(B)},
    \ee
which is a function of $\phi$ and the elements of $E_d$.
We are now ready to write down the unbarred metric in terms of barred quantities.
Written explicitly,
    \begin{align}
    g_\mn&=\fr{1}{f_0}\bra{\bar{g}_\mn-f_1\phi_\mu\phi_\nu-2\phi_{(\mu}\xi_{\nu)}} \nonumber \\
    &=\fr{1}{f_0}\brb{\bar{g}_\mn-\bra{f_1+2\Xi}\phi_\mu\phi_\nu-2\sum_{n=1}^{d-1}\sum_{m=1}^{n}\sum_{A,B}f_{n+1}^{(A)}J_{n m}^{(AB)}\phi_{(\mu}\bar{\Ds}_{\nu)}\bar{\chi}_m^{(B)}},
    \end{align}
where each $\chi_n^{(A)}$ on the right-hand side should be regarded as a function of $\bar{\chi}_m^{(B)}$ with $m\le n$.
This provides the inverse disformal transformation associated with Eq.~\eqref{GDTd}.

\subsection{Generalized disformal theories}\label{ssec:GDtheories}

So far, we have clarified the invertibility condition for generalized disformal transformations of the form~\eqref{GDTd} that involve arbitrary higher-order covariant derivatives of the scalar field [see Eq.~\eqref{inv_cond}].
By performing such an invertible transformation on Horndeski/U-DHOST theories, one can further extend the framework of GDH/GDU theories.\footnote{If one starts from a theory where the scalar field is nondynamical (e.g., the cuscuton~\cite{Afshordi:2006ad} or its extension~\cite{Iyonaga:2018vnu,Iyonaga:2020bmm}), then the scalar field remains nondynamical in the resultant theory.}
Schematically, the action of Horndeski/U-DHOST theories has the following form:
    \be
    S_{\rm g}[g,\phi]=\int \D^4x\sqrt{-g}\,\mL(g_\mn,R_{\mu\nu\lambda\sigma},\phi,\phi_\mu,\phi_\mn),
    \label{seed_action}
    \ee
with $R_{\mu\nu\lambda\sigma}$ being the four-dimensional Riemann tensor and $\phi_\mn\coloneqq \na_\mu\na_\nu\phi$.
Note that Eq.~\eqref{S_g-3D} can always be recast in this form.
The generalized disformal transformation of the action~\eqref{seed_action} is obtained by replacing $g_\mn$ with $\bar{g}_\mn[g,\phi]$.
More concretely, we perform the following replacements:
    \be
    \begin{split}
    \sqrt{-g}&\to\sqrt{-g}\,f_0\mF^{1/2}, \\
    R^\mu{}_{\nu\la\sigma}&\to R^\mu{}_{\nu\la\sigma}+2\na_{[\la} C^\mu{}_{\si]\nu}+2C^\mu{}_{\alpha[\la}C^\alpha{}_{\si]\nu}, \\
    \phi_\mn&\to \phi_\mn-C^\la{}_\mn\phi_\la.
    \end{split}
    \ee
Here, we have defined
    \be
    C^\la{}_\mn\coloneqq \bar{g}^{\la\alpha}\bra{\na_{(\mu}\bar{g}_{\nu)\alpha} -\fr{1}{2}\na_\alpha\bar{g}_\mn}, \label{Ctensor}
    \ee
which corresponds to the change of the Christoffel symbol under the generalized disformal transformation.
Note that $\phi$ and $\phi_\mu$ remain unchanged under the transformation.
We emphasize that our generalized disformal transformation~\eqref{GDTd} respects the properties~[\ref{propertyA}] and [\ref{propertyB}], and hence the generalized disformal theories described by the action~$\ti{S}_{\rm g}[g,\phi]\coloneqq S_{\rm g}[\bar{g},\phi]$ accommodate consistent matter coupling.

Finally, let us briefly comment on the relation to the so-called effective field theory (EFT) of inflation/dark energy~\cite{Creminelli:2006xe,Cheung:2007st,Gubitosi:2012hu}.
When one studies cosmology based on scalar-tensor theories, the EFT description provides a useful and robust framework for studying the dynamics of perturbations.\footnote{The EFT was recently extended to accommodate vector-tensor theories~\cite{Aoki:2021wew} and solids/fluids~\cite{Aoki:2022ipw}, where the symmetry breaking pattern is different from the one in the case of scalar-tensor theories.}
In this context, one assumes that the background scalar field has a timelike profile so that one can choose the unitary gauge where the scalar DOF is eaten by the metric.
Such a scalar field spontaneously breaks the time diffeomorphism and the residual spacetime symmetries are only the spatial diffeomorphisms.
Therefore, in the unitary gauge, the action is written in terms of geometrical quantities that respect spatial covariance as well as those respecting full spacetime covariance.
It is straightforward to write down the action of cosmological perturbations up to the leading order in the derivative expansion, and it accommodates the case of Horndeski theories, as it should.
The EFT was extended to incorporate the DHOST theories in Ref.~\cite{Langlois:2017mxy} and then further extended to incorporate the GDH/GDU theories in Ref.~\cite{Takahashi:2023jro}.
The idea of Ref.~\cite{Takahashi:2023jro} is to start from the EFT action up to the leading order in the derivative expansion and then perform the generalized disformal transformation~\eqref{GDT2} that involves covariant derivatives of the scalar field up to the second order.
In particular, it turned out that the effects of the GDH/GDU theories appear already at the level of linear perturbations (i.e., at the level of quadratic action).
The same idea can be applied to the transformation~\eqref{GDTd} that involves arbitrary higher-order covariant derivatives of the scalar field.
However, in this case, the effects of the new terms would show up only at the level of second- or higher-order perturbations (i.e., at the level of cubic- or higher-order action).
For instance, the first two new building blocks of the transformation~\eqref{GDTd}, $\mU=\Ds_\alpha X\Ds^\alpha\mZ$ and $\mV=\Ds_\alpha\mZ\Ds^\alpha\mZ$, are at least cubic and quartic order in perturbations, respectively.
(Note that $\mZ$ is defined by $\mZ=\Ds_\alpha X\Ds^\alpha X$ and $\Ds_\mu$ corresponds to the spatially covariant derivative under the unitary gauge, and hence $\mZ$ starts at the quadratic order.)
It should be noted that the above discussion assumes a homogeneous and isotropic cosmological background, and the situation could change if one considers an inhomogeneous background.
Recently, the EFT of perturbations on an arbitrary background spacetime with a timelike scalar profile was formulated in Ref.~\cite{Mukohyama:2022enj} and then applied to black hole perturbations in Refs.~\cite{Mukohyama:2022skk,Mukohyama:2023xyf,Konoplya:2023ppx}.
It would be intriguing to extend this EFT to incorporate our generalized disformal theories, which we leave for future work.

\section{Conclusions}\label{sec:conc}

Invertible disformal transformations provide a useful tool to investigate ghost-free scalar-tensor theories.
Recently, a general class of invertible disformal transformations with covariant derivatives of the scalar field up to the second order was constructed in Ref.~\cite{Takahashi:2021ttd}, which enabled us to construct a novel class of ghost-free scalar-tensor theories, i.e., GDH/GDU theories.
These generalized disformal theories lie outside the conventional DHOST class and hence would exhibit novel phenomena that can be tested with experiments/observations.
The consistency of matter coupling was studied in Refs.~\cite{Takahashi:2022mew,Naruko:2022vuh,Takahashi:2022ctx,Ikeda:2023ntu}, and it was shown that there exists a nontrivial subclass of generalized disformal theories where matter fields can be coupled without introducing an unwanted extra DOF in the unitary gauge.
As we discussed in \S\ref{sec:2nd}, there are two properties (i.e., the properties~[\ref{propertyA}] and [\ref{propertyB}]) that ensure the consistency of matter coupling.

Along this line of thought, in the present paper, we have constructed a class of invertible disformal transformations with arbitrary higher-order covariant derivatives of the scalar field, respecting the properties~[\ref{propertyA}] and [\ref{propertyB}].
The explicit form of the transformation has been obtained in \S\ref{ssec:trnsf_law} [see Eq.~\eqref{GDTd}], and the invertibility condition has been studied in \S\ref{ssec:inv_cond}.
We have shown that the invertibility condition can be written in the simple form of Eq.~\eqref{inv_cond}.
Moreover, we have specified the independent functional DOFs that characterize the invertible subclass of the generalized disformal transformation [see the discussion around Eq.~\eqref{f0f1}].
As discussed in \S\ref{ssec:GDtheories}, these transformations can be used to further extend the framework of the GDH/GDU theories, keeping the consistency of matter coupling in the unitary gauge.

There are several possible future directions.
As mentioned in \S\ref{ssec:GDtheories}, it would be intriguing to extend the EFT of cosmological perturbations~\cite{Creminelli:2006xe,Cheung:2007st,Gubitosi:2012hu} or the EFT of perturbations on an arbitrary background spacetime with a timelike scalar profile~\cite{Mukohyama:2022enj} to incorporate generalized disformal theories with arbitrary higher-order derivatives.
It is also interesting to study the screening mechanism.
As shown in Refs.~\cite{DeFelice:2011th,Koyama:2013paa,Kase:2013uja}, the Vainshtein screening mechanism~\cite{Vainshtein:1972sx,Babichev:2013usa} is built-in in the Horndeski theories.
On the other hand, in the DH theories, the authors of Refs.~\cite{Kobayashi:2014ida,Koyama:2015oma,Saito:2015fza,Crisostomi:2017lbg,Langlois:2017dyl,Dima:2017pwp} showed that the Vainshtein screening is partially broken inside astrophysical bodies.
Since our generalized disformal theories involve arbitrary higher-order spatial derivatives, their effect on small scales could be much more significant than in known theories.
We leave these issues for future studies.


\acknowledgments{
The author was supported by Japan Society for the Promotion of Science KAKENHI Grant Nos.~JP22KJ1646 and JP23K13101.
}


\bibliographystyle{mybibstyle}
\bibliography{bib}

\end{document}